\documentclass{emulateapj}
\usepackage{natbib}
\usepackage{epsfig}
\usepackage{graphicx}
\usepackage{subfigure}
\usepackage{float}
\usepackage{amsmath}
\usepackage{color}
\usepackage{amssymb}
\usepackage{amsfonts}
\usepackage[colorlinks,linkcolor=blue,anchorcolor=green,citecolor=blue]{hyperref}
\usepackage{amssymb}
\usepackage{upgreek}
\usepackage{color}
\usepackage[normalem]{ulem}
\usepackage{verbatim}

\def\be{\begin{equation}}
\def\ee{\end{equation}}
\shorttitle{FRB 121102: A STAR QUAKE-INDUCED REPEATER?}
\shortauthors{Wang et al.}
\begin{document}

\title{FRB 121102: a star quake-induced repeater?}
\author{Weiyang Wang\altaffilmark{1,2,3}, Rui Luo\altaffilmark{4,5}, Han Yue\altaffilmark{6}, Xuelei Chen\altaffilmark{1,2,7}, Kejia Lee\altaffilmark{4,8}, Renxin Xu\altaffilmark{3,4,5}}
\affil{$^1$Key Laboratory for Computational Astrophysics, National Astronomical Observatories, Chinese Academy of Sciences, 20A Datun Road, Beijing 100012, China}
\affil{$^2$University of Chinese Academy of Sciences, Beijing 100049, China}
\affil{$^3$School of Physics and State Key Laboratory of Nuclear Physics and Technology, Peking University, Beijing 100871, China}
\affil{$^4$Kavli Institute for Astronomy and Astrophysics, Peking University, Beijing 100871, China}
\affil{$^5$Department of Astronomy, School of Physics, Peking University, Beijing 100871, China}
\affil{$^6$School of Earth and Space Science, Peking University, Beijing 100871, China}
\affil{$^7$Center for High Energy Physics, Peking University, Beijing 100871, China}
\affil{$^8$National Astronomical Observatories, Chinese Academy of Sciences,  Beijing 100012, China}
\email{wywang@bao.ac.cn}
\begin{abstract}
Since its initial discovery, the Fast radio burst (FRB) FRB 121102 has been found to be repeating with millisecond-duration pulses. Very recently, 14 new bursts were detected by the Green Bank Telescope (GBT) during its continuous monitoring observations. In this paper, we show that the burst energy distribution has a power law form which is very similar to the Gutenberg-Richter law of earthquakes. In addition, the distribution of burst waiting time can be described as a Poissonian or Gaussian distribution, which is consistent with earthquakes, while the aftershock sequence that exhibits some local correlations. These findings suggest that the repeating FRB pulses may originate from the starquakes of a pulsar. Noting that the soft gamma-ray repeaters (SGRs) also exhibit such distributions, the FRB could be powered by some starquake mechanisms associated with the SGRs, including crustal activity of a magnetar and solidification-induced stress of a new-born strangeon star. These conjectures could be tested with more repeating samples.
\end{abstract}

\keywords{pulsars: general - radiation mechanisms: non-thermal - radio continuum: general - stars: neutron}

\section{Introduction}
Fast Radio Bursts (FRBs) are mysterious millisecond-duration radio flashes with high flux densities and prominent dispersive features \citep{lorimer07,Keane12,thornton13,mas15,ravi16,Bannister17,Bhandari17,Caleb17}.
The observed large values of dispersion measure (DM) are in the range of $\sim100-2600\rm{\,pc\,cm^{-3}}$, which indicate that FRBs are probably of extragalactic or even cosmological origins (e.g., \citealt{Katz16a,sch16}).
These transient phenomena stimulate interests of astrophysicist significantly, especially FRB 121102 which is the only repeater that has been detected so far, with an estimated burst energy $\sim10^{37-38}$\,erg \citep{Spitler14,Spitler16}.
The optical counterpart of the repeater, has been identified as a host faint star-forming dwarf galaxy which is at a redshift of $z = 0.193$ \citep{chatterjee17,kok17,tendulkar17}.

A persistent radio source which is thought to be associated with the repeater, was identified at a distance of $\lesssim40\,$pc from the FRB location \citep{chatterjee17,marcote17}.
\cite{ofek17} also found 11 source candidates with luminosities of $\nu L_{\nu}> 3\times10^{37}\,\rm{erg\,s^{-1}}$ which are spatially associated with disks or star-forming regions of galaxies rather than be in galactic center.
The persistent radio source is likely to be a pulsar wind nebula \citep{2017ApJ...843L..26B,dai17,kas17}.
With an active pulsar producing bursts repeatedly, ejecta or ultra-relativistic electron/positron pair winds may sweep up and heat the nebula that emits synchrotron radio emissions.
Additionally, \cite{wax17} calculated some stringent constraints on the persistent source's age.
The local environment of FRB source would be tested by the variation of DM which has not shown significant evolution \citep{yang17}.

It is proposed that FRBs are highly likely to be associated with pulsars and more than a few efforts have been made to understand their origins.
For instance, FRB is supposed to result from a pulsar's magnetosphere suddenly combed by a nearby cosmic plasma stream \citep{zhang17}.
Also, \cite{dai16} suggested that the repeater is originated from a highly magnetized pulsar traveling through asteroid belts.
Alternatively, in a neutron star (NS)-white dwarf binary system, the accreted magnetized materials may trigger magnetic reconnection that is accounting for FRBs \citep{Gu16}.
Other possibility is that the radio emission is produced by the interaction between a highly relativistic flow and nebula which is powered by a new-born millisecond magnetar \citep{murase16,2017ApJ...843L..26B,dai17}. This process might couple with a long gamma-ray burst or an ultraluminous supernova \citep{Metzger17}.
FRB are also interpreted by the model of supergiant pulse or giant flare from young pulsar or magnetar \citep{pop10,kul14,cordes16,Katz16b}.

Very recently, 14 bursts above threshold of 10 sigma in two 30-minute scans were detected by Breakthrough Listen Digital Backend with the C-band receiver at the Green Bank Telescope \citep{ATel17}.
In this paper, we propose that this repeating burst may arise from a pulsar's starquake.
The burst energy and waiting time distributions as well as the time decaying of the seismicity rate, are shown in Section 2.
The scenarios of possible origins will be discussed in Section 3.
In Section 4, we make our conclusions.

\section{The Earthquake-like behaviors of FRB 121102}

\begin{table}
\begin{center}
\caption{14 bursts of FRB 121102 in continuous observations by Green Bank Telescope}
\begin{tabular}{ccc}
\hline \hline
No. & MHD & Energy Density\\ & & ($\rm{Jy\,\mu s}$)\\
\hline
1 & 57991.577788085 & 114.2\\
2 & 57991.580915232 & 24.8\\
3 & 57991.581342500 & 112.5\\
4 & 57991.581590370 & 61.0\\
5 & 57991.581720752 & 54.6\\
6 & 57991.584516806 & 144.5\\
7 & 57991.586200359 & 25.3\\
8 & 57991.586510463 & 27.7\\
9 & 57991.589595602 & 29.3\\
10 &	57991.590822338 & 26.5\\
11 &	57991.594435069 & 49.6\\
12 &	57991.599814375 & 32.4\\
13 &	57991.607200359 & 49.4\\
14 &	57991.616266551 & 25.7\\
\hline \hline
\label{tab1}
\end{tabular}
\end{center}

{\bf Notes.} Data are quoted from \cite{ATel17}, where the event 11E and 11F are actually a same burst  (see \citealt{Katz17b}, for a review of close burst pairs). The energy of this burst is calculated to the average value of 11E and 11F.
\end{table}

\begin{figure}
\includegraphics[width=0.48\textwidth]{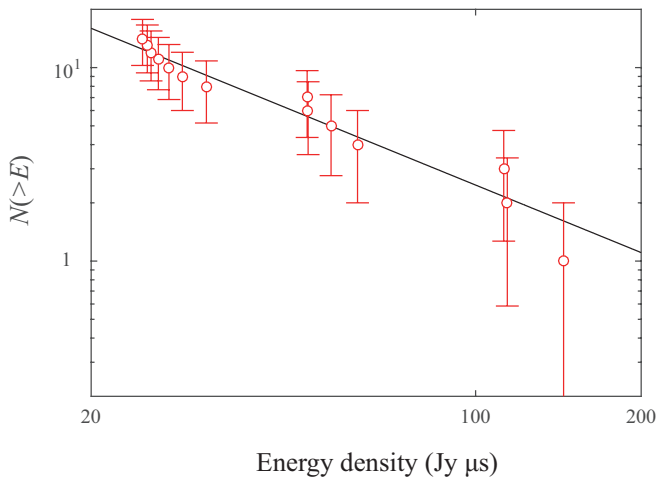}
\caption{\small{The cumulative distribution of each burst energy for FRB 121102, respectively. The solid black line is the best fitting power law of which index is $\alpha_{E}=1.16\pm0.24$ with $95\%$ confidence.}}
\label{fig1}
\end{figure}

\begin{figure}
\includegraphics[width=0.48\textwidth]{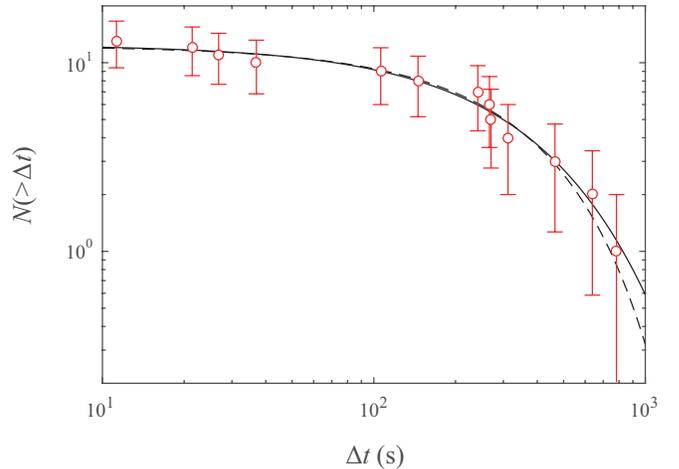}
\caption{\small{The cumulative distribution of waiting time for FRB 121102, respectively. The solid black line is the fitting curve with the Poissonian function where $\lambda=(3.05\pm0.48)\times10^{-3}\rm{s^{-1}}$, while the dashed black line is the fitting curve with Gaussian function where $\tau=(1.13\pm0.20)\times10^3$\,s and $\sigma=(1.03\pm0.16)\times10^{3}$\,s. Both fittings are derived for a $95\%$ confidence.}}
\label{fig2}
\end{figure}

\begin{figure}
\includegraphics[width=0.48\textwidth]{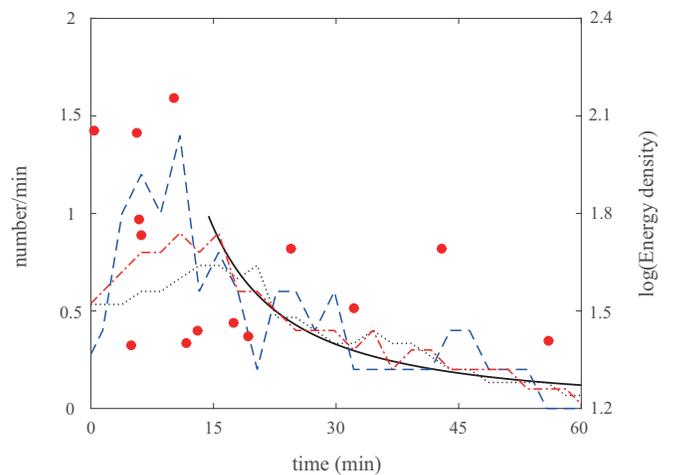}
\caption{\small{The time decaying of the seismicity rate (left axis) with different binning width of time, including 5\,min (blue dashed line), 10\,min (red dashed-dotted line) and 15\,min (black dotted line), comparing with magnitudes of burst (red dots, right axis). The bold solid black line is the best fitting of the aftershock sequences (from time$>14.4$\,min) by the Omori law, where $p=1.42\pm0.24$, $K=39.71\pm39.49$ and a fixed $C=1$\,min with $95\%$ confidence.}}
\label{fig3}
\end{figure}

Considering the unity of telescope selection and wavebands ($4-8$\,GHz), and the completeness of the continuous observations (290\,mins), here we adopt the parameters of repeating bursts from latest continuous monitoring GBT observations \cite{ATel17} rather than including the Arecibo events \citep{Spitler16} and previous GBT events \citep{sch16}.
Parameters of these 14 bursts are shown in Table \ref{tab1}.

We make statistics on the observational parameters and it turns out that  the burst rate as function of burst energy is consistent.
Considering the binning width can affect the fitting result, here we calculate the cumulative distribution to avoid this problem.
The burst energy is proportional to the observed energy density.
With a power law distribution for the number distribution of burst energy $N(E)\propto E^{-\alpha}$, the cumulative distribution can be obtained,
\begin{equation}\label{eq3}
N(>E)\propto \int^{\infty}_{E}E^{-\alpha}dE\propto E^{-\alpha+1}.
\end{equation}
The fluctuations of events for the cumulative distribution are assumed to follow a random statistic, $\sigma(E)=\sqrt{N(>E)}$.
The cumulative energy distribution of events for each burst energy (energy density) is well fitted by a power law with an index $\alpha_{E}=\alpha-1=1.16\pm0.24$, shown in Figure \ref{fig1}.
This power law distribution is consistent with the Gutenberg-Richter power law (i.e., $N(E)\propto E^{-2}$, \citealt{Gutenberg56}) which is a well-known earthquake law.

Furthermore, the statistics of waiting times contain much significant informations about occurrences and correlations of events.
The waiting time $\Delta t$ is defined as the interval time between the adjacent detected FRB events in the continuous monitoring observation.
For a simple Poisson process, the cumulative distribution of waiting time can be described by a simple exponential function,
\begin{equation}\label{eq2}
N(>\Delta t)\propto e^{-\lambda \Delta t},
\end{equation}
where $\lambda$ is the burst rate, which is constant.
Also, with an assumed Gaussian distribution of the number distribution $N(\Delta t)$, the cumulative distribution of the waiting time is,
\begin{equation}\label{eq3}
N(>\Delta t)\propto \int^{\infty}_{\Delta t}\mathrm{exp}^{-\frac{(\Delta t-\tau)^2}{\sigma^2}}d(\Delta t)\propto 1-\mathrm{erf}(\Delta t).
\end{equation}
As shown in Figure \ref{fig2}, the cumulative distribution of the waiting time are plotted, fitted by the exponential function with $\lambda=(3.05\pm0.48)\times10^{-3}\,\rm{s^{-1}}$ and equation (\ref{eq3}) with $\tau=(1.13\pm0.20)\times10^3$\,s and $\sigma=(1.03\pm0.16)\times10^{3}$\,s.
The waiting time distribution can be represented by a simple Poissonian or Gaussian distribution, which are extracted from statistic of earthquakes \citep{pep94,Leonard2001}.
There may be additional burst events with detection threshold of around 7 over full bandwidth of 4 GHz. However, they are not listed in \cite{ATel17} because they are relatively weak and need more analysis. These bursts might have narrow frequency spread and thus do not show high signal to noise ratio. This may not affect our model significantly while affect the fitting parameters.
Earthquakes from different regions are regarded as random processes of independent and uncorrelated events, while aftershocks which occur in shorter time intervals are correlated.

Additionally, combining with magnitudes of burst, the burst rates are plotted with different binning widths of time and fitted by a power law, shown in Figure \ref{fig3}.
For short-range temporal correlations between earthquakes, the time decaying of the seismicity rates of an aftershock sequence can be interpreted by an empirical relationship, i.e. Omori law \citep{omori94,Utsu61,Utsu95}.
By the Omori law, the seismicity rates decay with time, can be expressed by a power law,
\begin{equation}\label{omori}
n(t)=\frac{K}{(C+t)^p},
\end{equation}
where $n(t)$ is the seismicity rate and the time decaying rate of seismicity is controlled by the third constant $p$, which typically falls in the range of 0.9-1.5 \citep{Utsu95}.
For an aftershock sequence, the variation of $p$ may be controlled by the structural heterogeneity, stress and temperature \citep{Utsu95}.
We fixed the parameter $C$ at 1\,min, and then the number rate at $t=0$ is close to $K$ which is the peak value of the number rate curve.
Thus only one parameter $p$ which controls the decaying rate is estimated.
With the occurrence time of 14 FRB events, the burst rates can be well fitted by the power law determined a index $p$ value of $1.42\pm0.24$, which is consistent with the index for earthquakes.
Therefore, the bursts are followed by an earthquake-like aftershock sequence.

\section{Possible origins of FRB 121102}

As we see from Section 2, the FRB repeater exhibits several features commonly found in earthquakes.
In fact, nonlinear dissipative systems always show self-organized criticality (SOC) behaviors.
With a solid crust or stiff equation of state (EOS) of pulsar, a star can build up stresses that makes the crust cracks and adjust stellar shape to reduce its deformation.
The characters for these processes of statistical independence, nonlinear coherent growth and random rise times, are consistent with a SOC system \citep{SOC}.
According to the SOC theory, NS quakes are giant catastrophic events like earthquakes and probably accompanied by global seismic vibrations or oscillations.
Hence, starquakes in a pulsar, share similar statistical distributions with earthquakes.
Here we present two possible scenarios as following.

One scenario is that a normal NS with a solid crust and a superfluid core, and with a strong toroidal magnetic field (i.e., magnetar).
NS can form a solid crust quickly after its birth.
The stellar shape changes from oblate to spherical configuration, as well as thermal and dynamic responses, will induce stresses in the crusts.
When stress buildup in the dense solid crust beyond a yield point, a starquake would happen with sudden energy release that supports FRB.
This process also brings a slight change of moment of inertia with an abrupt jump of angular frequency that is the so-called pulsar glitch \citep{ruderman69,Link96}.
For a highly magnetized NS, the stellar crust is coupling with the magnetosphere.
In this scenario, starquakes induce the magnetic curl or twist ejected into the magnetosphere from crust in a few milliseconds \citep{Thompson02,Thompson17}.
Electrons in the magnetosphere are suddenly accelerated to ultra-relativistic velocity by magnetic reconnection \citep{zhang11}, move along the magnetic field lines, resulting in producing curvature radiation.
The characteristic frequency of the curvature radiation is
\begin{equation}\label{eq5}
\nu_{\mathrm c}=\frac{3c\gamma^3}{4\pi R_{\rm c}}=7.16\times(\frac{\gamma}{100})^3(\frac{10\,\rm km}{R_{\rm c}})\,\rm GHz,
\end{equation}
where $R_{\rm c}$ is the curvature radius with a typical value of $\sim10$\,km and $\gamma$ is the Lorentz factor of electrons.
With detected FRB frequency $\nu_{\rm c}=4-8$\,GHz, a $\gamma\approx50-100$ is required.
If generated by such stress-induced reconnections (e.g., \citealt{Lyutikov15}), the FRB has an estimated duration timescale,
\begin{equation}\label{eq6}
t_{\rm {rec}}\sim\frac{L}{v_{\rm A}}\sim1-10\,\rm ms,
\end{equation}
where the scale of the reconnection-unstable zone $L\sim 1-10$\,km, and the Alfv$\acute{\rm e}$n velocity is $v_{\rm A}\simeq B/(4\pi \rho_{\rm c})^{1/2}\sim0.01c$, in which $\rho_{\rm c}\simeq10^{14}\,\rm{g\,cm^{-3}}$ is the average mass density of the crust.

The sudden elastic and magnetic energy release in the crustal stress is estimated,
\begin{equation}\label{eq11}
\delta E_{\rm{cru}}=4\pi R^2h_{\rm c}\sigma\delta \varepsilon,
\end{equation}
where $h_{\rm c}\simeq R/30\simeq 0.3$\,km is the crustal thickness \citep{Thompson17}, $\varepsilon$ is the shear strain and the total stress including crustal shear stress and Maxwell stress in magnetosphere, is
\begin{equation}\label{eq12}
\sigma=\sqrt{(\mu\varepsilon)^2+(\frac{BB_z}{4\pi})^2},
\end{equation}
in which $\mu$ is the shear modulus, $B$ is the surface magnetic field and $B_z$ is the component of magnetic field perpendicular to the direction of plastic flow.
Within the crust, the force balance $\mu\varepsilon\simeq BB_z/(4\pi)$ implies that
\begin{equation}\label{eq13}
\delta E_{\rm{cru}}\simeq4.2\times10^{46}(\frac{BB_z}{10^{30}\,\rm{G^2}})(\frac{R}{10\,\rm{km}})^2\delta\varepsilon\,\rm{erg}.
\end{equation}
$\delta\varepsilon$ is smaller than $\sim10^{-2}$ (e.g., \citealt{hof12}).
Here, the energy release can meet energy requirements of FRBs while not for soft gamma repeaters (SGRs).
However, a plastic flow can be initiated when the elastic crustal deformation exceeds a critical value, launching a thermo-plastic wave that dissipates the magnetic energy inside the crust \citep{Bel14}.
This mechanism might bring much more energy from the inner crust in which it stores a SGR-required magnetic energy of $\gtrsim10^{47}$\,erg with a interior magnetic field of $\sim10^{16}$\,G \citep{Lander16}.
The Ohmic dissipation in this process can be neglected because of the long timescale \citep{2014MNRAS.445.2777F}.
The timescale for the local energy release is $t_{\rm tw}\sim4\pi\eta/BB_{z}$, where $\eta$ is the viscosity.
If the energy releases quickly enough for FRB, a viscosity of $\sim10^{26}\,\rm{erg\,s\,cm^{-3}}$ is required.
In addition, the transition to hydromagnetic instability of the magnetar core may offer larger energy \citep{Thompson17} supporting SGRs.

The other scenario is that a new-born strangeon star (SS), which has a stiff EOS \citep{lai17a} and could release more elastic energy than that of the solid crust of a normal NS.
At early age, SS may shrink its volume abruptly by solidification-induced stress, and a bulk-variable starquake happens.
Basically, there are two kinds of quakes in a solid star: bulk-invariable (type I) and bulk-variable (type II) starquake \citep{zhou04,zhou14}.
Here, type II starquake is more likely to be dominant because the elastic energy accumulation of type I quake is not sufficient to produce such short time interval quakes in this quake sequence (see equation (39) in \citealt{zhou14} for a test of $t=100$\,s).
The elastic and gravitational energy release during type II starquake is
\begin{equation}\label{eq4}
\begin{split}
\delta E_{\rm{g}}=\frac{3GM^2}{5R}\frac{\delta R}{R}=\frac{3GM^2}{10R}\frac{\delta \Omega}{\Omega}\\
\simeq10^{53}(\frac{M}{1.4M_{\odot}})^2(\frac{10\,\mathrm{km}}{R})(\frac{\delta\Omega}{\Omega})\,\mathrm{erg},
\end{split}
\end{equation}
where $G$ is the gravitational constant, $M$ is the stellar mass, $\delta R/R$ is the strain and $\delta \Omega/\Omega$ is the amplitude of a glitch (e.g., $10^{-9}-10^{-6}$, \citealt{Alpar94,Alpar96}).
This energy is large enough to support a FRB and possibly associating with SGR.
Bulk-variable starquakes are accompanied by the change of electrostatic energy \citep{Katz17a} and some electrodynamic activities in magnetosphere.
A giant quake can power energetic relativistic outflow to produce the observed prompt emission of short-hard GRBs, and some aftershocks result in following X-ray flares observed \citep{xu06}.
Starquakes may also lead to the magnetic reconnection that accelerates electrons, and these charges move along the magnetic field lines, emitting curvature radiation.

The duration timescale of the magnetic reconnection in this scenario can be obtained from equation (\ref{eq6}).
Besides, these short time interval quakes might be motivated by an initial shock which is type I quake dominate.
The waiting time of the initial shock can be obtained \citep{zhou04},
\begin{equation}\label{eq7}
t_{\rm i}=\frac{\sigma_{\rm c}}{\dot\sigma},
\end{equation}
where $\sigma_{\rm c}$ is the critical stress and $\dot\sigma$ can be denoted as,
\begin{equation}\label{eq8}
\dot\sigma=\frac{3\pi I\dot P}{R^3P^3}\approx 9.42\times10^{27}(\frac{\dot P}{1\,\rm{s\,s^{-1}}})(\frac{1\,\rm{s}}{P})^3,
\end{equation}
in which $I\approx10^{45}\,\rm{erg\,s^2}$ is the moment of inertia, $P$ is the rotation period and $\dot P$ is period derivative of the star.
Then, the waiting time of the initial shock can be written as,
\begin{equation}\label{eq9}
\log({\frac{t_{\rm i}}{1\,\rm s}})=\log({\frac{\sigma_{\rm c}}{1\,\rm erg\,cm^{-3}}})+3\log({\frac{P}{1\,\rm s}})-\log({\frac{\dot P}{1\,\rm s\,s^{-1}}})-28.
\end{equation}
The rotation period $P\sim10$\,ms for a new-born rapidly rotating SS.
From equation (\ref{eq9}), the critical stress is estimated to $10^{19-22}\,\rm{erg\,cm^{-3}}$ which is consistent with strangeon stars \citep{zhou04}.

Starquakes are magnetically powered in a NS while elastically and gravitationally powered in a SS.
In these scenarios, the toroidal oscillation, which might be derived from starquakes \citep{Bas07}, propagates into the magnetosphere and changes its charge density that brings an induced electric potential \citep{lin15}.
The electric potential \citep{ruderman75,chen93} is estimated,
\begin{equation}\label{V}
\Delta V\simeq2.1\times10^{12}(\frac{\Omega_{\rm{osc}}}{10\,\rm{kHz}})^{1/7}(\frac{B}{10^{14}\,\rm{G}})^{-1/7}(\frac{R}{10\,\rm{km}})^{4/7}\,\rm V,
\end{equation}
where stellar oscillation frequency is estimated as $\Omega_{\rm {osc}}\sim c/R\sim30$\,kHz that enlarges the size of radio beam.
Within this picture, a magnetar is most likely to produce electron/positron pair plasma.
The electron/positron pair plasma production due to the electric potential is the necessary condition for coherent radio emission.
The potential enhances voltage along the gap which accelerates electrons to higher Lorentz factors emitting curvature radiation.
Then, the pulsar becomes radio loud (i.e., beyond the pulsar death line) until oscillations damp out in which the magnetosphere becomes inactive and radio emissions evaporate.
Therefore, FRB may be the ``oscillation'' of a dead pulsar at near pulsar death line.

\section{Conclusion and dicussion}
Basically, we found that the behaviors of the repeating FRB 121102 are earthquake-like.
The distribution of burst energy exhibits a Gutenberg-Richter power law form which is a well-known earthquake distribution.
And the distribution of waiting time, can be characterized as a Poissonian or Gaussian distribution, which are consistent with earthquakes as well as the local correlated aftershock sequence.
The possible origins of the repeater are discussed including crustal activity of a magnetar and solidification-induced stress of a new-born SS.
Both possible origins might be associated with SGRs which are difficult to detect at cosmological distance.
Statistic distributions of burst energy and duration time show that FRB 121102 is very similar to SGR $1806-20$ \citep{JCAP17}.
Also, SGR $1806-20$ share some distinctive properties with earthquake that indicates SGRs are indeed powered by starquakes \citep{cheng96}, and the giant flares of SGRs are suggested to be quake-induced \citep{xu06}.

In addition, the observed continuous bursts with the modeled occurrence rate $\lambda=11.0\,\rm{hr}^{-1}$ for Poissonian, while $\tau^{-1}=3.2\,\rm{hr}^{-1}$ for Gaussian, may originate from some uncorrelated quakes.
In that case, these quakes are suggested foreshocks, storing energy and motivate a main quake.
Then an aftershock sequence, which may be caused by some local coherent deformations before a new equilibrium sets up, occurs.
The motivated shocks are non-Poissonian and not rotation-powered dominate, while the type I starquake may lead the initial shock begins when the stresses exceed a certain threshold.
Hence, the next repeating FRB might be waiting for $\sim10^6$\,s because a long time to store elastic energy is needed \citep{zhou14}.

And, the latest FRB volumetric rate including all of repeating bursts is calculated as $R_{\rm FRB}\sim10^{-5}\,\rm{Mpc^{-3}\,yr^{-1}}$ out to redshift of 1 \citep{law17}.
A pulsar, which has a solid crust or stiff EOS, would be natural to have glitches as the result of starquakes.
From statistics of pulsar glitches, the number of glitches per year can be interpreted as \citep{esp11},
\begin{equation}\label{glitch}
\dot N_{\rm g}\simeq0.003\times(\frac{\dot \nu}{10^{-15}\,\rm{Hz\,s^{-1}}})^{0.47}\,\rm{yr^{-1}},
\end{equation}
where $\dot \nu$ is rotational frequency derivative.
For a typical millisecond pulsar with $P\sim10$\,ms and $\dot P\sim10^{-21}\,\rm{s\,s^{-1}}$, the number of glitches per year can be evaluate to $\sim3\times10^{-4}\,\rm{yr^{-1}}$.
A total number of glitches $N_{\rm g}\sim3\times10^3$ can be estimated during the pulsar life-time $\sim10$\,Myr.
It is hypothesized that the hydrogen-poor superluminous supernovae (SLSNe-I) are powered by millisecond magnetars.
The volumetric birth rate of SLSNe-I is $R_{\rm {SLSN}}=10^{-8}\,\rm{Mpc^{-3}\,yr^{-1}}$ \citep{gal12}.
Therefore, we estimate the FRB volumetric rate $R_{\rm {FRB}}\simeq N_{\rm g}R_{\rm {SLSN}}\sim3\times10^{-5}\,\rm{Mpc^{-3}\,yr^{-1}}$.
This inferred event rate from FRB/SLSNe-I associated events is consistent with the observational FRB events.

Starquakes associating with some X-ray or Gamma-ray bursts in a normal NS share similar behaviors with that in a SS.
While the X-ray spectra might be different in these scenarios.
In a SS atmosphere, thermal X-rays from the lower layer of a normal NS atmosphere are prohibited, relatively more optical/UV photons and a energy cutoff at X-ray bands are exhibited \citep{2017arXiv170503763W}.
Considering a NS at 1\,Gpc with $2-8$\,keV flux of $\sim2\times10^{-16}\,\rm{erg\,cm^{-2}\,s^{-1}}$ which is consistent with the X-ray limit of {\it Chandra} and {\it XMM-Newton},  the luminosity is calculated to $\sim10^{39}\,\rm{erg\,s^{-1}}$.
Such distant source is too faint to be detected by current X-ray telescopes except with a supper-Eddington luminosity.
The normal NS and SS have different EOS that are most likely to be tested by gravitational wave and electromagnetic radiation from mergers of compact stars further \citep{GW,Lai17b}.

We expect to detect more repeating events. More constraints on the mysterious origin of FRB will be given by the statistics growing samples. The earthquake-like behaviors, including distributions of energy and waiting time for the repeater, are expected to be tested by more continuous monitoring observations.

\acknowledgments
We are grateful to Stephen Justham at National Astronomical Observatories, Chinese Academy of Sciences, for discussions. This work is supported by National Key R\&D Program of China (No. 2017YFA0402602), the National Natural Science Foundation of China (11373030, 11673002, 11633004 and U1531243), Frontier Science Key Project (QYZDJ-SSW-SLH017) of Chinese Academy of Sciences, and the Ministry of Science and Technology (2016YFE0100300).


\begin{thebibliography}{}
\expandafter\ifx\csname natexlab\endcsname\relax\def\natexlab#1{#1}\fi

\bibitem[Abbott(2017)]{GW} Abbott B. P., et al., 2017, Phys. Rev. Lett., 119, 161101

\bibitem[Alpar \& Baykal(1994)]{Alpar94} Alpar, M.~A., \& Baykal, A.\ 1994, \mnras, 269, 849

\bibitem[Alpar et al.(1996)]{Alpar96} Alpar, M.~A., Chau, H.~F., Cheng, K.~S., \& Pines, D.\ 1996, \apj, 459, 706

\bibitem[Aschwanden(2011)]{SOC} Aschwanden, M.~J.\ 2011, Self-Organized Criticality in Astrophysics, by Markus J.~Aschwanden.~ Springer-Praxis, Berlin ISBN 978-3-642-15000-5, 416p.,

\bibitem[Bannister et al.(2017)]{Bannister17} Bannister, K.~W., Shannon, R.~M., Macquart, J.-P., et al.\ 2017, \apjl, 841, L12

\bibitem[Bastrukov et al.(2007)]{Bas07} Bastrukov, S.~I., Chang, H.-K., Takata, J., Chen, G.-T., \& Molodtsova, I.~V.\ 2007, \mnras, 382, 849

\bibitem[Beloborodov \& Levin(2014)]{Bel14} Beloborodov, A.~M., \& Levin, Y.\ 2014, \apjl, 794, L24

\bibitem[Beloborodov(2017)]{2017ApJ...843L..26B} Beloborodov, A.~M.\ 2017, \apjl, 843, L26

\bibitem[Bhandari et al.(2017)]{Bhandari17} Bhandari, S., Keane, E.~F., Barr, E.~D., et al.\ 2017, arXiv:1711.08110

\bibitem[Caleb et al.(2017)]{Caleb17} Caleb, M., Flynn, C., Bailes, M., et al.\ 2017, \mnras, 468, 3746

\bibitem[Chatterjee et al.(2017)]{chatterjee17} Chatterjee, S., Law, C.~J., Wharton, R.~S., et al.\ 2017, \nat, 541, 58

\bibitem[Chen \& Ruderman (1993)]{chen93} Chen, K, \& Ruderman, M. A. \ 1993, \apj, 402, 264

\bibitem[Cheng et al.(1996)]{cheng96} Cheng, B., Epstein, R.~I., Guyer, R.~A., \& Young, A.~C.\ 1996, \nat, 382, 518

\bibitem[Cordes et al.(2016)]{cordes16} Cordes, J.~M., \& Wasserman, I.\ 2016, \mnras, 457, 232

\bibitem[Dai et al.(2016)]{dai16} Dai, Z.~G., Wang, J.~S., Wu, X.~F., \& Huang, Y.~F.\ 2016, \apj, 829, 27

\bibitem[Dai et al.(2017)]{dai17} Dai, Z.~G., Wang, J.~S., \& Yu, Y.~W.\ 2017, \apjl, 838, L7

\bibitem[Espinoza et al.(2011)]{esp11} Espinoza, C.~M., Lyne, A.~G., Stappers, B.~W., \& Kramer, M.\ 2011, \mnras, 414, 1679

\bibitem[Fujisawa \& Kisaka(2014)]{2014MNRAS.445.2777F} Fujisawa, K., \& Kisaka, S.\ 2014, \mnras, 445, 2777

\bibitem[Gajjar et al.(2017)]{ATel17} Gajjar, V.,, Siemion, A.~P.~V., MacMahon, D.~H.~E., et al.\ 2017, The Astronomer's Telegram, 10675

\bibitem[Gal-Yam(2012)]{gal12} Gal-Yam, A.\ 2012, Science, 337, 927

\bibitem[Gu et al.(2016)]{Gu16} Gu, W.-M., Dong, Y.-Z., Liu, T., Ma, R., \& Wang, J.\ 2016, \apjl, 823, L28

\bibitem[Gutenberg \& Richter(1956)]{Gutenberg56} Gutenberg, B., \& Richter, C. F. 1956, Bull. seiem. Soc. Am, 46, 105

\bibitem[Haensel et al.(2007)]{hae07} Haensel, P., Potekhin, A.~Y., \& Yakovlev, D.~G.\ 2007, Astrophysics and Space Science Library, 326, Haensel

\bibitem[Hoffman \& Heyl(2012)]{hof12} Hoffman, K., \& Heyl, J.\ 2012, \mnras, 426, 2404

\bibitem[Kashiyama \& Murase(2017)]{kas17} Kashiyama, K., \& Murase, K.\ 2017, \apjl, 839, L3

\bibitem[Katz(2016a)]{Katz16a}Katz, J. I. 2016a, \apj, 818, 19

\bibitem[Katz(2016b)]{Katz16b} Katz, J.~I.\ 2016b, \apj, 826, 226

\bibitem[Katz(2017a)]{Katz17a} Katz, J.~I.\ 2017a, \mnras, 469, L39

\bibitem[Katz(2017b)]{Katz17b} Katz, J.~I.\ 2017b, arXiv:1708.07234

\bibitem[Keane et al.(2012)]{Keane12}Keane, E. F., Stappers, B. W., Kramer, M., \& Lyne, A. G. 2012, \mnras, 425, L71

\bibitem[Kokubo et al.(2017)]{kok17} Kokubo, M., Mitsuda, K., Sugai, H., et al.\ 2017, \apj, 844, 95

\bibitem[Kulkarni(2014)]{kul14} Kulkarni, S. R., Ofek, E. O., Neill, J. D., Zheng, Z., \& Juric, M. 2014, \apj, 797, 70

\bibitem[Law et al.(2017)]{law17} Law, C.~J., Abruzzo, M.~W., Bassa, C.~G., et al.\ 2017, arXiv:1705.07553

\bibitem[Lai \& Xu(2017)]{lai17a} Lai X. Y. \& Xu R. X., 2017, Journal of Physics: Conf. Series 861, 012027. arXiv: 1701.08463

\bibitem[Lai et al.(2017)]{Lai17b} Lai, X.~Y., Yu, Y.~W., Zhou, E.~P., Li, Y.~Y., \& Xu, R.~X.\ 2017, arXiv:1710.04964

\bibitem[Lander et al.(2015)]{lander15} Lander, S.~K., Andersson, N., Antonopoulou, D., \& Watts, A.~L.\ 2015, \mnras, 449, 2047

\bibitem[Lander(2016)]{Lander16} Lander, S.~K.\ 2016, \apjl, 824, L2

\bibitem[Leonard et al.(2001)]{Leonard2001} Leonard, T., Papasouliotis, O., \& Main, I.~G.\ 2001, \jgr, 106, 13

\bibitem[Lin et al.(2015)]{lin15} Lin, M.-X., Xu, R.-X., \& Zhang, B.\ 2015, \apj, 799, 152

\bibitem[Link \& Epstein(1996)]{Link96} Link, B., \& Epstein, R.~I.\ 1996, \apj, 457, 844

\bibitem[{{Lorimer} {et~al.}(2007){Lorimer}, {Bailes}, {McLaughlin},
  {Narkevic}, \& {Crawford}}]{lorimer07}
{Lorimer}, D.~R., {Bailes}, M., {McLaughlin}, M.~A., {Narkevic}, D.~J., \&
  {Crawford}, F. 2007, Science, 318, 777

\bibitem[Lyutikov(2015)]{Lyutikov15} Lyutikov, M. 2015, \mnras, 447, 1407

 \bibitem[Masui et al.(2015)]{mas15} Masui, K., Lin, H.-H., Sievers, J., et al.\ 2015, \nat, 528, 523

\bibitem[{{Marcote} {et~al.}(2017){Marcote}, {Paragi}, {Hessels}, {Keimpema},
  {van Langevelde}, {Huang}, {Bassa}, {Bogdanov}, {Bower}, {Burke-Spolaor},
  {Butler}, {Campbell}, {Chatterjee}, {Cordes}, {Demorest}, {Garrett}, {Ghosh},
  {Kaspi}, {Law}, {Lazio}, {McLaughlin}, {Ransom}, {Salter}, {Scholz},
  {Seymour}, {Siemion}, {Spitler}, {Tendulkar}, \& {Wharton}}]{marcote17}
{Marcote}, B., {Paragi}, Z., {Hessels}, J.~W.~T., {et~al.} 2017, \apjl, 834, L8

\bibitem[Metzger et al.(2017)]{Metzger17} Metzger, B.~D., Berger, E., \& Margalit, B.\ 2017, \apj, 841, 14

\bibitem[Mitra \& Deshpande(1999)]{beam} Mitra, D., \& Deshpande, A.~A.\ 1999, \aap, 346, 906

\bibitem[{{Murase} {et~al.}(2016){Murase}, {Kashiyama}, \&
  {M{\'e}sz{\'a}ros}}]{murase16}
{Murase}, K., {Kashiyama}, K., \& {M{\'e}sz{\'a}ros}, P. 2016, \mnras, 461, 1498

\bibitem[Ofek(2017)]{ofek17} Ofek, E. O. 2017, \apj, 846, 44

\bibitem[Omori(1894)]{omori94} Omori, F. 1894, Journal of the College of Science, Imperial University of Tokyo. 7, 111

\bibitem[Pepke(1994)]{pep94} Pepke, S. L., Calson, J. M., \& Shaw, B. E. 1994, J. geophys. Revs, 99, 6769

\bibitem[Popov \& Postnov(2010)]{pop10} Popov, S.~B., \& Postnov, K.~A.\ 2010, Evolution of Cosmic Objects through their Physical Activity, 129

\bibitem[Ravi et al.(2016)]{ravi16} Ravi, V., Shannon, R.~M., Bailes, M., et al.\ 2016, Science, 354, 1249

\bibitem[Ruderman(1969)]{ruderman69} Ruderman, M.\ 1969, \nat, 223, 597

\bibitem[Ruderman \& Sutherland(1975)]{ruderman75} Ruderman, M.~A., \& Sutherland, P.~G.\ 1975, \apj, 196, 51

\bibitem[Scholz et al.(2016)]{sch16} Scholz, P., Spitler, L.~G., Hessels, J.~W.~T., et al.\ 2016, \apj, 833, 177

\bibitem[Spitler(2014)]{Spitler14} Spitler, L. G., Cordes, J. M., Hessels, J. W. T., et al. 2014, \apj, 790, 101

\bibitem[Spitler(2016)]{Spitler16} Spitler, L. G., Scholz, P., Hessels, J. W. T., et al. 2016, \nat, 531, 202

\bibitem[{{Tendulkar} {et~al.}(2017){Tendulkar}, {Bassa}, {Cordes}, {Bower},
  {Law}, {Chatterjee}, {Adams}, {Bogdanov}, {Burke-Spolaor}, {Butler},
  {Demorest}, {Hessels}, {Kaspi}, {Lazio}, {Maddox}, {Marcote}, {McLaughlin},
  {Paragi}, {Ransom}, {Scholz}, {Seymour}, {Spitler}, {van Langevelde}, \&
  {Wharton}}]{tendulkar17}
{Tendulkar}, S.~P., {Bassa}, C.~G., {Cordes}, J.~M., {et~al.} 2017, \apjl, 834, L7

\bibitem[Thompson et al.(2002)]{Thompson02} Thompson, C., Lyutikov, M., \& Kulkarni, S. R. 2002, \apj, 574, 332

\bibitem[Thompson et al.(2017)]{Thompson17} Thompson, C., Yang, H., \& Ortiz, N.\ 2017, \apj, 841, 54

\bibitem[{{Thornton} {et~al.}(2013){Thornton}, {Stappers}, {Bailes},
  {Barsdell}, {Bates}, {Bhat}, {Burgay}, {Burke-Spolaor}, {Champion}, {Coster},
  {D'Amico}, {Jameson}, {Johnston}, {Keith}, {Kramer}, {Levin}, {Milia}, {Ng},
  {Possenti}, \& {van Straten}}]{thornton13}
{Thornton}, D., {Stappers}, B., {Bailes}, M., {et~al.} 2013, Science, 341, 53

\bibitem[Utsu(1961)]{Utsu61} Utsu, T. 1961, Geophysical Magazine, 30, 521

\bibitem[Utsu et al.(1995)]{Utsu95} Utsu, T., Ogata, Y., Matsu'ura, R. S. 1995, Journal of Physics of the Earth, 43, 1

\bibitem[Wang et al.(2017)]{2017arXiv170503763W} Wang, W., Feng, Y., Lai, X., et al.\ 2017, arXiv:1705.03763

\bibitem[Wang \& Yu(2017)]{JCAP17} Wang, F.~Y., \& Yu, H.\ 2017, JCAP, 3, 023

\bibitem[Weltevrede et al.(2011)]{Weltevrede11} Weltevrede, P., Johnston, S., \& Espinoza, C.~M.\ 2011, \mnras, 411, 1917

\bibitem[Waxman(2017)]{wax17} Waxman, E.\ 2017, \apj, 842, 34

\bibitem[Xu et al.(2006)]{xu06} Xu, R.~X., Tao, D.~J., \& Yang, Y.\ 2006, \mnras, 373, L85

\bibitem[Yang \& Zhang(2017)]{yang17} Yang, Y. P., \& Zhang, B.\ 2017, \apj, 847, 22

\bibitem[Zhang \& Yan(2011)]{zhang11} Zhang, B., \& Yan, H. 2011, \apj, 726, 90

\bibitem[Zhang (2017)]{zhang17} Zhang, B. 2017, \apj, 836, L32

\bibitem[Zhou et al.(2004)]{zhou04} Zhou, A.~Z., Xu, R.~X., Wu, X.~J., \& Wang, N.\ 2004, Astroparticle Physics, 22, 73

\bibitem[Zhou et al.(2014)]{zhou14} Zhou, E.~P., Lu, J.~G., Tong, H., \& Xu, R.~X.\ 2014, \mnras, 443, 2705


\end{thebibliography}
\end{document}